\documentclass[lettersize,journal]{IEEEtran}
\usepackage{amsmath,amsfonts}
\usepackage{bbm}
\usepackage{bbold}

\usepackage{svg}
\usepackage{svg-extract}
\usepackage{array}
\usepackage{caption}

\usepackage{textcomp}
\usepackage{stfloats}
\usepackage{url}
\usepackage{verbatim}
\usepackage{graphicx}
\usepackage{subfigure}
\usepackage[ruled]{algorithm2e}
\usepackage{algorithmic}

\def\BibTeX{{\rm B\kern-.05em{\sc i\kern-.025em b}\kern-.08em
    T\kern-.1667em\lower.7ex\hbox{E}\kern-.125emX}}
\usepackage{balance}
\begin{document}
\title{Minimizing Sensor Allocation Cost 
for Crowdsensing On-street Parking Availability}
\author{
\IEEEauthorblockN{Boyu Pang,  Ruizhi Liao  and Yinyu Ye}
\thanks{This work was supported in part by NSFC (61902332), Shenzhen STIC (JCYJ20180508162604311), Longgang STIB (20200030) and CUHK(SZ) URA. (Corresponding author: Ruizhi Liao). 

Boyu Pang is with School of Science and Engineering, The Chinese University of Hong Kong, Shenzhen, China (e-mail: boyupang@link.cuhk.edu.cn).

Ruizhi Liao is with Shenzhen Key Laboratory of IoT Intelligent Systems and Wireless Network Technology, China and School of Humanities and Social Science, The Chinese University of Hong Kong, Shenzhen, China  (e-mail: rzliao@cuhk.edu.cn).

Yinyu Ye is with Department of Management Science \& Engineering, Stanford University  (e-mail: yinyu-ye@stanford.edu).}}

\maketitle

\begin{abstract}
    In recent years, innovative roadside parking vacancy crowdsensing solutions have emerged as a cost-effective alternative to traditional methods, which can significantly reduce sensor installation and maintenance expenses. This  crowdsensing scheme relies on  vehicles equipped with sensors, such as buses and taxis, roaming around urban streets to detect on-street parking availability. Therefore, the accuracy of this scheme strongly depends on the vehicles' routes and the frequency of their passage through  parking spots. This paper presents an integer programming-based optimal sensor allocation model to ensure the detection accuracy of the scheme while using the minimum number of sensing kits or probing vehicles. Moreover, a customized heuristic algorithm is proposed to hasten the solution  process. 
    Numerical simulations using the street dataset from San Francisco confirm the model's ability to reduce probing vehicle usage while ensuring detection accuracy. Thus, our approach represents an effective means of optimizing roadside parking detection in a crowdsensing way.
    
\end{abstract}


\section{Introduction}
\IEEEPARstart{S}{eeking} an unoccupied on-street parking space in urban areas is a time-consuming task for drivers. Moreover, the act of street cruising while seeking parking intensifies the congestion in turn. A downtown traffic study \cite{congestion1} highlights that cruising vehicles looking for on-street parking spaces can account for up to 30 percent of total traffic flows, potentially leading to further congestion. As modern metropolises continue to urbanize and traffic networks become increasingly complex, smart parking resolutions of such issues become imperative.

Parking guidance and information (PGI) system \cite{pgi} is a typical solution to the city parking problem. A PGI system primarily serves two purposes: 1. providing on-street parking occupancy information; 2. suggesting drivers with the best parking spots, based on practical factors (e.g., weather and timing) or certain decision models. 

The common parking monitoring techniques employ fixed-sensing devices such as cameras, magnetic, infrared, ultrasonic, or radar sensors to gather real-time parking occupancy data \cite{fix1, fix2, fix3}.  The fixed-sensing systems have shown great potential in addressing urban parking pains, but monitoring on-street parking by fixed sensors is costly for two reasons: 1. Since the aggregate number of parking spots in a whole city is large, installing fixed sensors and relevant supporting facilities require exorbitant labor work. 2. Fixed sensors are exposed outdoors and thus vulnerable to damage from the ambient environment. Thus, the maintenance cost of fixed sensors can be significant.

\begin{figure*}[t] 
\centering 
\includegraphics[width=1\textwidth]{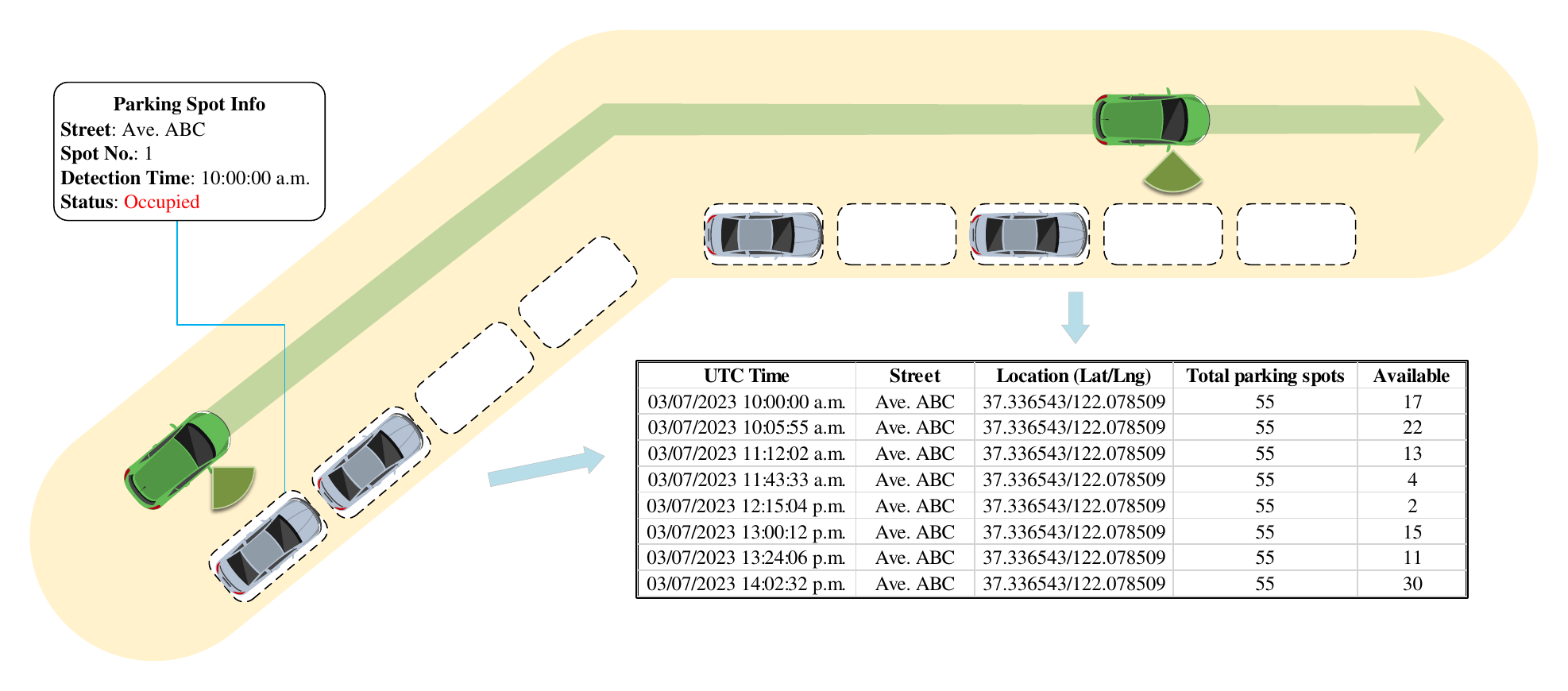} 
\caption{The working scenario of the crowdsensing system in detecting on-street parking availability} 
\label{Fig. work scenario} 
\end{figure*}

To overcome the high cost of fixed-sensing solutions while keeping high accuracy, Mathur et al. \cite{mobile0} proposed a mobile-sensing system as an effective  alternative (Fig. \ref{Fig. work scenario}). The proposed system was later echoed by Roman et al. \cite{mobile1} who incorporated a supervised learning algorithm in parking space recognition. The mobile-sensing system utilizes taxis or buses as sensing vehicles. Equipped with an ultrasonic sensing unit and Global Positioning System (GPS), a sensing vehicle is able to detect the nearby on-street parking spots along its route. Since each probing vehicle is able to detect multiple parking spots, the mobile-sensing system does not require assigning one sensing device to each individual parking spot. Therefore, the mobile-sensing system can provide accurate on-street parking information with much fewer  sensor requirements compared with fixed-sensing schemes. 


Insufficient probing vehicles will result in large time intervals between consecutive parking observations, leading to reduced accuracy of the system. Therefore, it is crucial to determine the minimum number of probing vehicles required for accurate parking availability detection. Bock et al. in \cite{vehicles num} conducted an empirical experiment in San Francisco, and demonstrated that employing 300 taxis as probing vehicles can achieve accuracy levels comparable to those of traditional fixed-sensing solutions. However, taxi trajectories are  unpredictable, and it is economically impossible to conduct such city-wide experiments  in every city. Consequently, the required minimum number of sensing vehicles  remains unclear. Thus, there is a need for a robust and scalable decision model that can precisely generate the minimum sensor units required for allocation, which can be universally applied to cities with diverse settings.

The main contributions of this paper are as follows. Firstly, we develop a mathematical optimization model that determines the minimum number of sensing vehicles required to detect roadside parking spaces in a crowdsourcing manner, resulting in a  50\% reduction in required sensing kits without sacrificing detection accuracy. The model is also highly scalable and flexible, allowing it to be easily applied to various city scenarios. Additionally, we introduce the STCB algorithm to efficiently solve the large-scale model. To evaluate the effectiveness and scalability of our approach, we use parking data from San Francisco \cite{parking data san Fran}. The numerical results demonstrate that the proposed algorithm can bring more than 100\% speedup during the solving process.

The remainder of this paper is organized as follows: Section \ref{sec: prob desc} introduces the mathematical optimization as well as its formulation. Section \ref{algo design} gives a detailed description of the proposed algorithm. Section \ref{num result} presents the numerical results. Section \ref{conclu} concludes the paper and discusses future  work.

\begin{figure*}[t] 
\centering 
\includegraphics[width=0.75\textwidth]{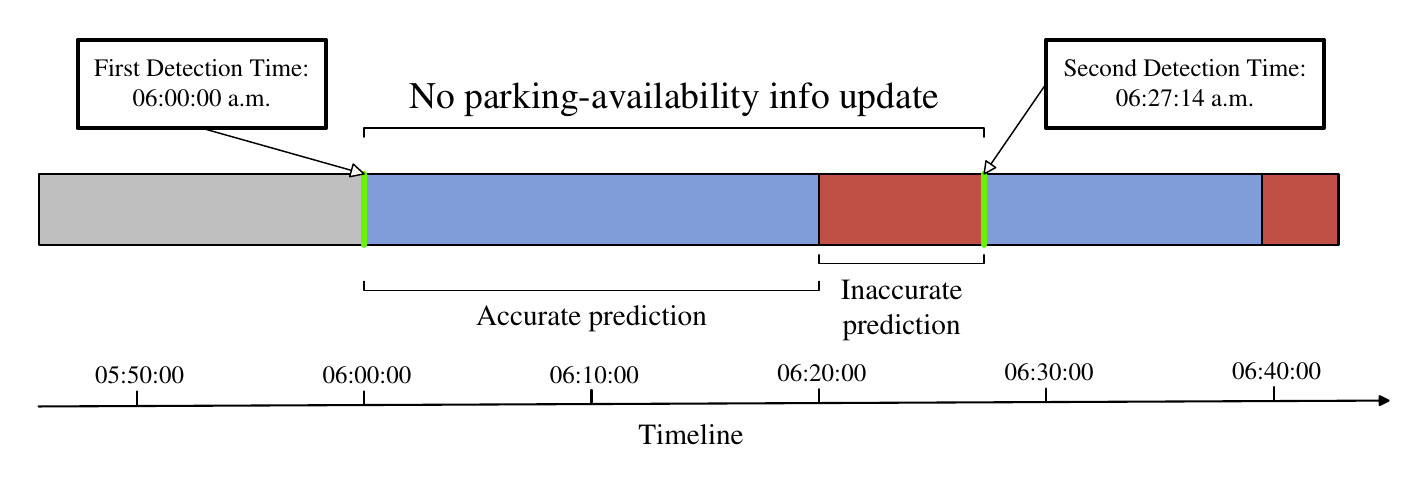} 
\caption{The timeline of two consecutive detections} 
\label{Fig. timeline} 
\end{figure*}

\section{Problem Formulation and  Optimization} \label{sec: prob desc}
\noindent 
This section provides a detailed description of the problem, including the definition of a uni-cost set-covering problem (SCP), and how to use a uni-cost SCP to develop a mathematical optimization model that determines the optimal number of sensors such that the time between two consecutive detections does not surpass a threshold.

Unlike the fixed-sensing solution, the mobile-sensing system cannot continuously update real-time information of on-street parking availability. Only when a probing vehicle passes by an on-street parking spot, the availability of that spot can be updated. If parking changes occurred between two consecutive detections, the parking availability information is inaccurate (Fig. \ref{Fig. timeline}). Thus, controlling the frequency between two consecutive scans is crucial to improve detection accuracy.

We assume that the routes and schedules of all buses, as well as the geolocations of all on-street parking spaces, can be obtained from open data platforms provided by cities. The objective of our study is to select the minimum number of buses to serve as probing vehicles while ensuring that the required  updating frequency is maintained. The parking information updating frequency is defined as follows.

\noindent \textbf{Definition 1:} For a given time interval of $T$ minutes and a given street,  the \textbf{parking information  updating frequency} is defined as the total number of probing vehicles that pass this street (thus detect parking spots on the street) during this time interval. The \textbf{minimum required updating frequency} is greater or equal to $1$ detection per $T$ minutes to ensure that the parking information for the street is updated at a sufficient frequency.


\subsection{Defining and Choosing Time Intervals} \label{def time-interval}
\noindent For all streets within the city, a minimum parking information updating frequency ($\ge1$ per $T$ minutes) is required during periods of high parking activity. In other words, we aim to ensure that the time duration between two consecutive detections does not exceed $T$ minutes during periods of high parking activity.

\begin{figure}[!h] 
\centering 
\includegraphics[width=0.5\textwidth]{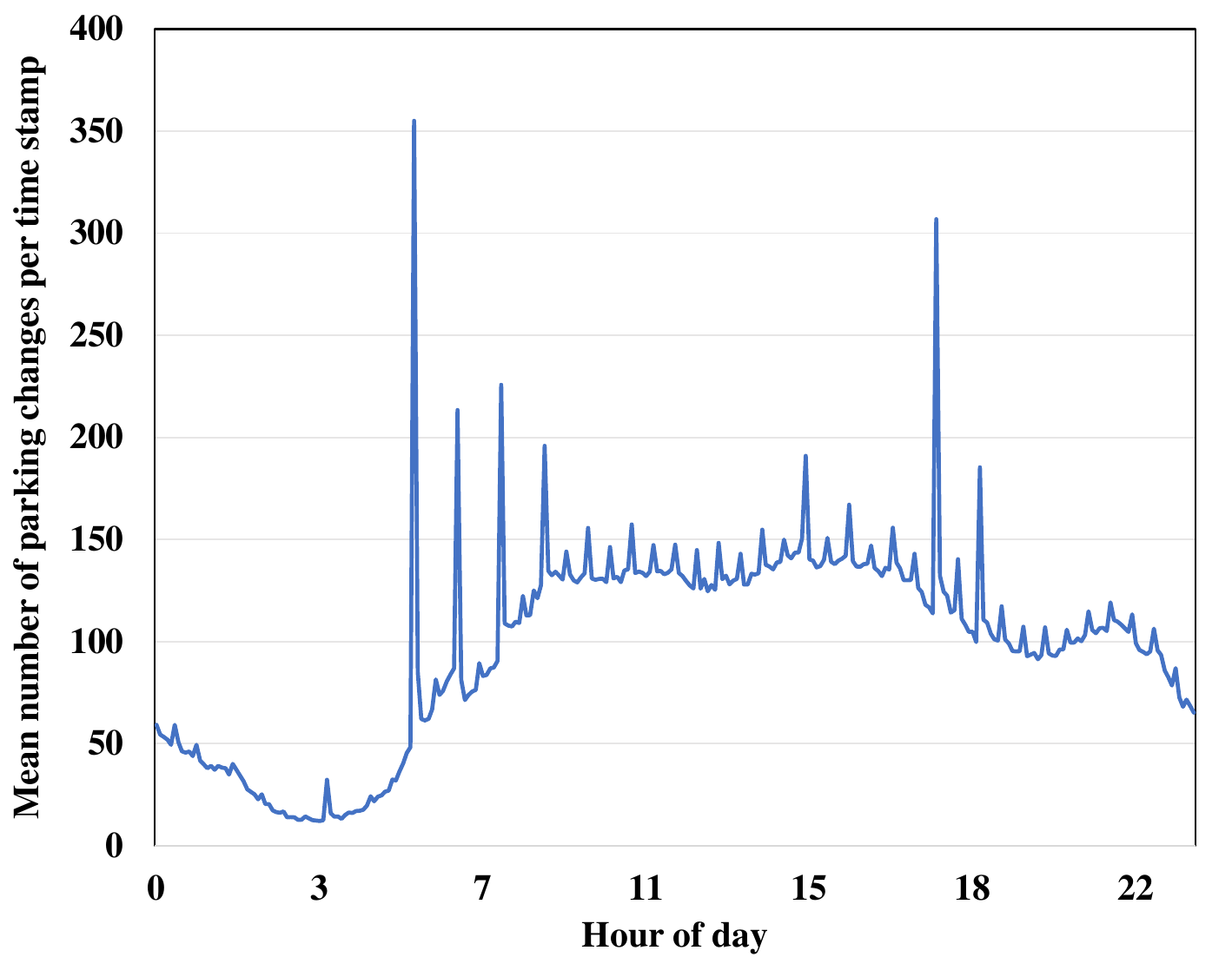} 
\caption{Parking fluctuations  on 420 streets in San Franciso} 
\label{Fig.parking change} 
\end{figure}

Utilizing historical on-street parking data from a city, we can determine the start and end times of high parking activity periods, either through direct estimation or by employing statistical rules. For example, Fig. \ref{Fig.parking change} illustrates the average parking  changes of 420 streets in San Francisco, spanning from June 13, 2013, to July 24, 2013 \cite{parking data san Fran}. It reveals that the high parking activity period in San Francisco commences around 6:00 and ends around 19:00.

\begin{figure}[!h]
    \centering
    \includegraphics[width=0.505\textwidth]{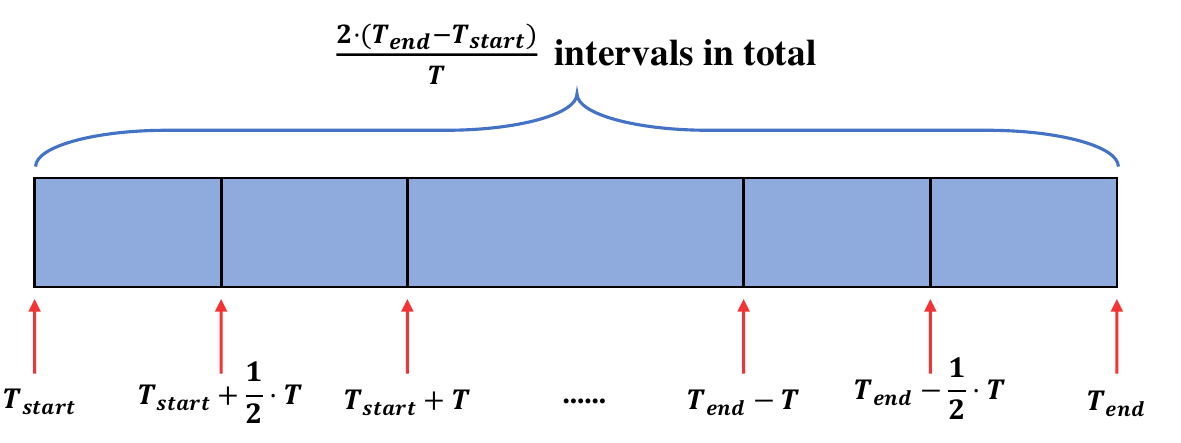}
    \caption{Dividing the high parking activity period into  intervals}
    \label{time intervals}
\end{figure}

Given that the high parking activity period starts at $T_{start}$ and ends at $T_{end}$, the whole period is divided into $\frac{2\cdot (T_{end}-T_{start})}{T}$ intervals, each with a duration of $\frac{1}{2}T$ minutes (Fig. \ref{time intervals}). Then we have the following proposition.

\noindent \textbf{Proposition 1:} If at least one detection happens in each time interval (duration=$\frac{1}{2}T$), then the minimum required updating frequency (i.e., $\geq1 $ detection per T min) is guaranteed.  

\noindent \textbf{Proof:} Assume a detection happens at time $t_a$, and the subsequent detection takes place at time $t_b$. There are two possible cases: 1. both detections happen in the same time interval; 2. the two detections are situated in two consecutive time intervals. Then, for both cases, $t_b-t_a \leq T$. Thus, it is ensured that at least one detection occurs in any T-minute time period, which signifies that the minimum required updating frequency is satisfied as per the definition.

\subsection{Trajectory Estimation} \label{subsec: prob cars}
\noindent In \cite{vehicles num}, taxis were selected as the probing vehicles. However, taxi trajectories are unpredictable, making them difficult to be tracked consistently. To address this issue, this study opts for buses as probing vehicles due to their  fixed routes and stable cruising speeds. For a given bus, with its route, schedule, and average cruising speed known, we can estimate whether it will pass by a specific on-street parking spot during a particular time interval. By iterating this process for every bus and every street, we can generate comprehensive trajectory information of a city. The pseudo-code for this process is provided below.

\begin{algorithm}[!htbp]
\caption{Initializing Trajectories}
\label{trajectory info algo}
    \begin{algorithmic}[1]
    \REQUIRE \thickspace \medspace route and schedule info for  bus $\mathbf{i}$ ($\mathbf{i=1, 2, ..., n}$), \\ \qquad location of  street $\mathbf{j}$ ($\mathbf{j=1, 2, ..., p}$), \\ \qquad
    every time interval $\mathbf{t}$ ($\mathbf{t=1, 2, ..., q}$)
    \ENSURE $\mathbf{N_t^j}$ (the set of buses that pass by street $\mathbf{j}$ \\ \qquad during the time  $\mathbf{t}$) for each $\mathbf{j}$ and $\mathbf{t}$
    \FOR{($\mathbf{i=1, 2, ..., n}$)}
        \FOR{($\mathbf{t=1, 2, ..., q}$)}
           \STATE $\mathbf{S_{streets}} \gets$ Compute streets that bus $\mathbf{i}$ passes \\ \qquad \qquad \quad by during the time $\mathbf{t}$
           \FOR{$\mathbf{j \in S_{streets}}$}
                \STATE $\mathbf{N_t^j}$.add($\mathbf{i}$)
            \ENDFOR
        \ENDFOR
    \ENDFOR
    \end{algorithmic}
\end{algorithm}

\subsection{Uni-Cost Set-Covering Formulation} \label{opt formula}

\noindent Previously, we defined the minimum required updating frequency and time intervals, as well as the initialization process for the trajectory sets  $\mathbf{N_t^j}$. Building on this foundation, we can formulate this problem into a uni-cost set-covering problem \cite{scp review}. Consider the following mathematical optimization problem:

\begin{equation} \label{unicost SCP}
\begin{aligned}
& \underset{\mathbf{x_i}}{\min}
& & \sum_{\mathbf{i=1}}^{\mathbf{n}} \mathbf{x_i} & \text{(Original Problem)} \\
& \textrm{s.t.}
& & \mathbf{\sum_{k \in N_t^{j=1}}}\mathbf{x_k} \ge \mathbf{1}, & \mathbf{\forall t=1, 2, ..., q} \\
&
& & ......\\
& 
& & \mathbf{\sum_{k \in N_t^{j=p}}}\mathbf{x_k} \ge \mathbf{1}, & \mathbf{\forall t=1, 2, ..., q} \\
& 
& & \mathbf{x_i} \in \mathbf{\{0, 1\}}, & \mathbf{\forall \space i=1, 2, ..., n}
\end{aligned}
\end{equation}
\noindent where $\mathbf{x_i}$'s are binary variables denoting whether to use bus $\mathbf{i}$ as probing vehicles (i.e., install a sensor on bus $\mathbf{i}$), and each constraint means that street $\mathbf{j}$ requires at least one detection during the time interval $\mathbf{t}$. For simplicity, this model can be reformulated more compactly as a typical uni-cost set-covering problem (Uni-Cost SCP):
\begin{equation} \label{compact SCP}
\begin{aligned}
& \underset{\mathbf{x}}{\min}
& & \mathbb{1}^\top \mathbf{x} & \text{(Uni-Cost SCP)} \\
& \textrm{s.t.}
& & \mathbf{A}\mathbf{x} \ge \mathbb{1} &\\
& 
& & \mathbf{x} \in \mathbf{\{0, 1\}^n }  
\end{aligned}
\end{equation}
where $\mathbf{A} \in \mathbb{R}^{\mathbf{m \times n}}$ is the constraint matrix ($\mathbf{m=p\cdot q}$).

\noindent \textbf{Remark 1:} $\mathbf{A}$ is a large-scale, sparse matrix with entries of either 0 or 1. Although the structure of this optimization formulation is simple, the scale of the problem is substantial  (e.g., a model with 420 streets and 52 time intervals as parameters already require 21840 constraints). Given that uni-cost SCPs are NP-hard in  mathematical optimization, solving such large-scale models is extremely challenging. Besides, the current heuristic algorithms for solving such problems are mostly problem-specific and even size/sparsity-specific. Therefore, we need to design an efficient algorithm that can provide a high-quality solution for this model.

\section{The Self-Trained Cardinality-Branching Algorithm} \label{algo design}

\begin{figure}[!b] 
\centering 
\includegraphics[width=0.5\textwidth]{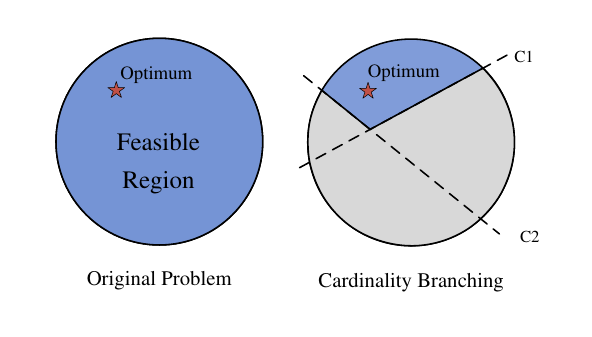} 
\caption{Cardinality-branching: a pair of cutting planes that contract the feasible region} 
\label{Fig. c-b} 
\end{figure}

\noindent SCPs are well-known as typical NP-hard integer programming problems. In other words, the difficulty of solving such problems increases exponentially with the size or density of the problem. Leveraging the specific structure of the constraint matrix, we develop a Self-Trained Cardinality-Branching (STCB) algorithm to enhance the solving speed.

\begin{figure}[!t] 
\centering 
\includegraphics[width=0.55\textwidth]{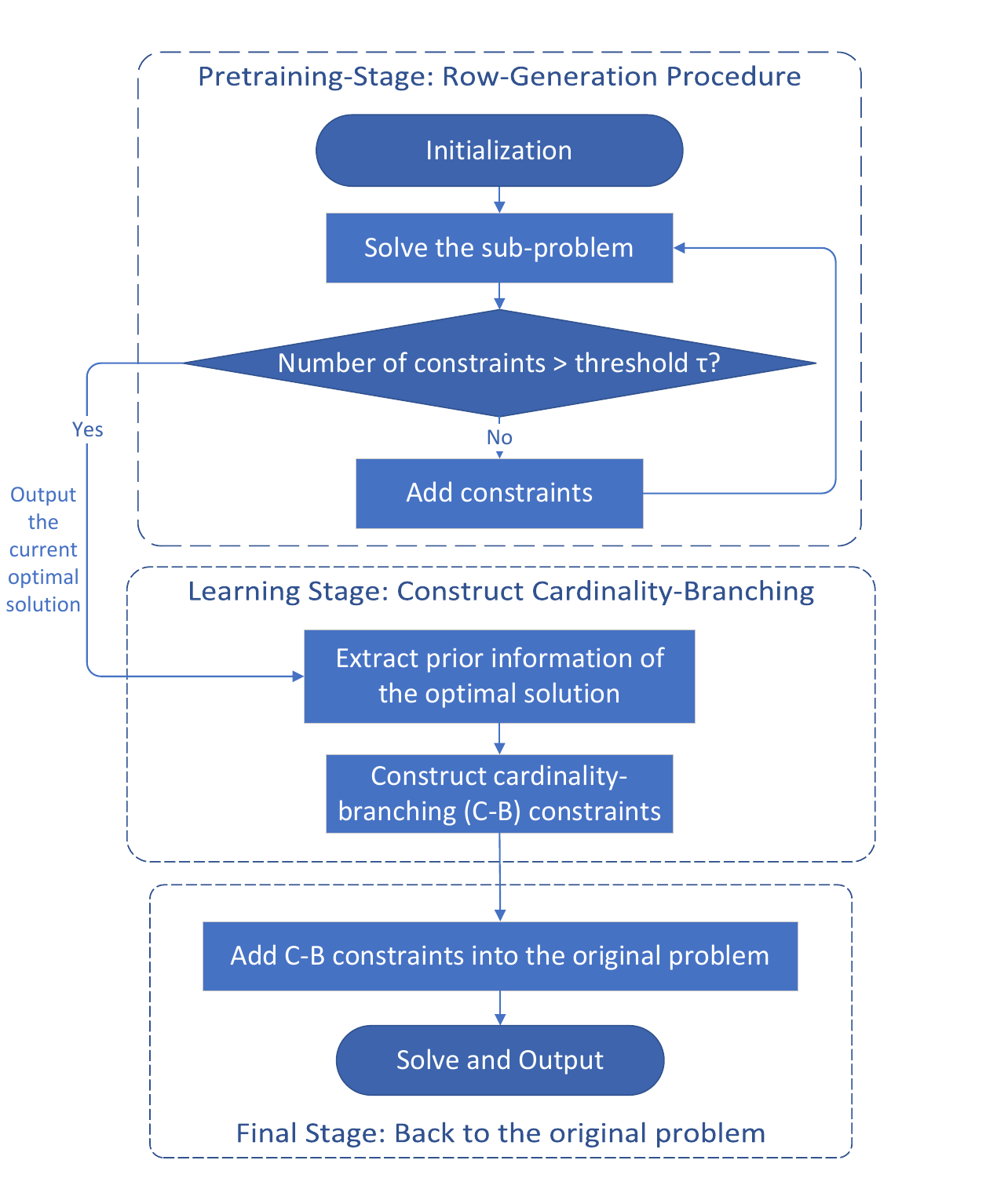} 
\caption{Flowchart of the STCB algorithm} 
\label{Fig. flowchart STCB} 
\end{figure}

The proposed algorithm is a "learn-then-optimize" method that combines the traditional row-generation approach with a cardinality-branching heuristic concept \cite{PreMIO}. The cardinality-branching heuristic scheme utilizes prior knowledge (historical data)  to generate a pair of cutting-plane constraints that significantly reduce the feasible region without compromising the optimal solution (Fig. \ref{Fig. c-b}). By contracting the feasible region, the search time for an optimal solution is shortened, as the algorithm no longer needs to explore redundant regions.

However, a critical issue persists: how can high-quality prior information be obtained without  significant time costs? In this study, we propose an STCB algorithm that uses the row-generation process for generating prior knowledge. In contrast to the classical row-generation algorithm, which aims to solve the original problem, the proposed STCB algorithm focuses on utilizing the first few iterations of the row-generation process to gather insights into the behavior of optimal solutions. This nature makes row generation  suitable for obtaining high-quality prior information. The flowchart of the STCB algorithm is depicted in Fig. \ref{Fig. flowchart STCB}.

The detailed explanations of row generation, cardinality-branching, and hyper-parameters learning, are presented in Section \ref{subsec: Row-gen}, \ref{subsec: C-B}, and \ref{subsec: hyper-para learn}, respectively.


\subsection{Pre-training Stage: Row Generation} \label{subsec: Row-gen}
\noindent Row-generation and column-generation methods are two classical approaches utilized in solving large-scale integer programming problems, such as SCPs \cite{row-gen method}. The underlying principle of both methods is to eliminate redundant parts of the constraint matrix and solve smaller sub-problems. Among the two methods, row generation is particularly suitable for cases where the number of constraints greatly exceeds the number of binary variables, resulting in a "tall" constraint matrix $\mathbf{A}$. In this study, the total number of constraints is equal to the product of  "total time intervals" and "total streets", which is typically much larger than the total number of buses. Furthermore, it is common for a street to be covered by a substantial number of bus routes, implying that the constraints associated with this street can be easily satisfied. Thus, the row-generation technique is primarily employed in this research.

In the row-generation method, we compute the sub-problem whose constraint set is a subset of the original problem's constraints by eliminating redundant rows. If the optimal solution of the sub-problem, $\mathbf{x^*_{sub}}$, is feasible for the original problem, then $\mathbf{x^*_{sub}}$ is also optimal for the original problem. However, a drawback of the row-generation method is that as the number of rows increases during iterations, the solving time in each iteration may grow rapidly, which significantly reduces the overall solving speed. Because the purpose of the pre-training stage is to generate prior knowledge of the optimal solution's behavior for the original problem, it is important to avoid spending excessive effort in this stage. To achieve this goal, we run the row-generation algorithm for only a limited number of iterations.


\noindent \textbf{Iterative R-G Algorithm:}

Consider a sub-problem (suppose the number of rows in $\mathbf{\tilde{A}}$ and $\mathbf{A}$ equals $\mathbf{\tilde{m}}$ and $\mathbf{m}$, respectively):

\begin{equation} \label{sub-problem}
\begin{aligned}
& \underset{\mathbf{x}}{\min}
& & \mathbb{1}^\top \mathbf{x} & \text{(sub-problem)} \\
& \textrm{s.t.}
& & \mathbf{\tilde{A}}\mathbf{x} \ge \mathbb{1} &\\
& 
& & \mathbf{x} \in \{0, 1\}^{n'} 
\end{aligned}
\end{equation}

\begin{itemize}
        \item Step 1: Start with no constraint ($\mathbf{\tilde{A}} = \mathbb{0}$).
        \item Step 2: Solve this sub-problem and generate the optimal solution $\mathbf{x^*_{sub}}$.
        \item Step 3: If $\mathbf{\tilde{m} \geq \tau}$, stop. Otherwise, Go to Step 4.
        \item Step 4: Find all violated constraints (rows) of the original problem. Pick $\mathbf{x}$ rows with the lowest density (fewest non-zero entries) and add them to the sub-problem. Go to Step 2.
    \end{itemize}
\textbf{Remark2}: In Step 4, the rationale for adding the lowest-density rows is that they are more challenging  than high-density rows.
For example, consider a situation where only 2 buses can pass by street $s_1$ at the time interval $t_1$ (resulting in a row with only 2 non-zero entries), while 100 buses can pass by street  $s_2$ at the time interval $t_2$ (yielding 100 non-zero entries). In this case, the second constraint is more likely to be satisfied from a probabilistic perspective.

\noindent \textbf{Remark3}: In practice, it is common for some streets, such as those in central areas, to be covered by several bus routes. Then, the cover constraints corresponding to these streets may be redundant. In such cases, even the optimal solution for a significantly smaller sub-problem might be able to satisfy the majority of the constraints in the original problem. As a result, the $\mathbf{x^*_{sub}}$ generated by the row-generation process can provide sufficient information for cardinality-branching constraints.

\subsection{Cardinality-Branching} \label{subsec: C-B}
\noindent The authors in \cite{PreMIO} introduced cardinality-branching and provided theoretical support using statistical theories. In essence, cardinality-branching constructs two branching hyper-planes that eliminate the redundant portions of feasible regions, guided by prior knowledge. The formula is presented as follows:

\begin{equation}
    \begin{aligned}
    \mathbf{\sum_{i \in S^+} x_i \ge \xi^+} \\
     \mathbf{\sum_{j \in S^-} x_j \le \xi^-}\\
    \mathbf{x_i, x_j \in \{0, 1\}}
    \end{aligned}
\end{equation}

The hyper-parameters are $S^+$, $S^-$, $\xi^+$, and $\xi^-$. The first two determine the division of variables into two sets, while the latter two determine the behavior of these variables. With the prior knowledge obtained from the row-generation process (Section \ref{subsec: Row-gen}), we can learn these four hyper-parameters using the method described in Section \ref{subsec: hyper-para learn}. To expedite the solving process of the original problem in Section \ref{opt formula}, we can incorporate the two constraints of cardinality-branching into the original problem. By eliminating redundant feasible regions, these two constraints reduce the search space and thus increase the solution speed.

\subsection{Hyper-parameters learning}
\label{subsec: hyper-para learn}
Selecting hyper-parameters for cardinality-branching is the crucial process that determines the effectiveness and accuracy of the cutting plane. This study proposes a method that extracts information from both the original constraint matrix $\mathbf{A}$ and the optimal solution $\mathbf{x^*_{sub}}$ of the sub-problem derived from the pre-training stage.

We cluster the decision variables $x_i$'s by conducting spectral clustering on the constraints of the original problem. Specifically, we construct a graph from these constraints and utilize the graph Laplacian to perform spectral clustering. The adjacency matrix of the graph is constructed using a Gram matrix of the constraint vectors, which is a matrix of inner products between the constraint vectors \cite{spectral-clustering}. The clustering process for decision variables can identify groups of variables that are strongly interconnected within the optimization problem, providing insights into the problem's structure to facilitate the discovery of the optimal solution. The implementation of all procedures is carried out through Algorithm \ref{Branch_by_clustering_constraints}.

\begin{algorithm}[!htbp]
        \caption{Clustering constraints}
        \label{Branch_by_clustering_constraints}
        \begin{algorithmic}[2]
            \REQUIRE constraint matrix $\mathbf{A}$, number of branches $k \in \mathbb{R}$, the optimal solution $\mathbf{x^*} \in \mathbb{R}^{n}$ of the sub-problem 
            
            \ENSURE $\{S^i\}, \{\xi^i\}$ (by default $k=2$, denote as $S^+, S^-, \xi^+, \xi^-$)

            \STATE $G \gets \mathbf{A}^\top \mathbf{A} \in \mathbb{R}^{n \times n}$
            \STATE $D \gets diag(G)$
            \STATE $\tilde{G} \gets D^{-1/2}G D^{-1/2}$ \#Obtain the normalized Gram matrix
            \STATE $\{S^i\} \gets $ Spectral Clustering(n\_clusters=$k$).fit($\tilde{G}$)
            \STATE $\{S^i\}_{i=1}^{k} \gets $ Sort $\{S^i\}$ by $\sum_{j\in S^i}\mathbf{x^*_j}$ \#Obtain $S^+, S^-$
            \STATE $\xi^{+} \gets \sum_{j \in S^+}\mathbf{x^*}_j$
            \STATE $\xi^{-} \gets |S^-|  -\sum_{j \in S^-}\mathbf{x^*}_j$
        \end{algorithmic}
\end{algorithm}

After selecting the hyper-parameters, we can add the cardinality-branching constraints into the original problem and solve it using any commercial solvers. The formulation is as follows:
\begin{equation} \label{compact SCP}
\begin{aligned}
& \underset{\mathbf{x}}{\min}
& & \mathbb{1}^\top \mathbf{x} &  \\
& \textrm{s.t.} 
& & \mathbf{A}\mathbf{x} \ge \mathbb{1} & \\
& 
& & \mathbf{\sum_{i \in S^+} x_i \ge \xi^+}  &\\
&
& & \mathbf{\sum_{j \in S^-} x_j \le \xi^-} &\\
&
& & \mathbf{x} \in \mathbf{\{0, 1\}^n } \\ 
\end{aligned}
\end{equation}
\section{Numerical Results}
\label{num result}
\noindent To validate the feasibility of the mathematical modeling and the performance of the algorithm, this study conducted a numerical experiment simulating the settings in San Francisco based on data from \cite{parking data san Fran}. The computational results demonstrate that the model proposed in Section \ref{algo design} can provide an efficient sensor allocation plan for the crowdsensing system. Compared to random allocation plans, such a deterministic solution can reduce the cost by half while maintaining the same detection accuracy level. Furthermore, this section also offers a performance evaluation of the proposed algorithm. 

The proposed algorithm was executed with an Intel Core i7-9750H CPU and 16GB RAM (12 logical processors, up to 12 threads). All optimization tasks were solved using the commercial solver - Gurobi 10.0.0.

\subsection{City Scenario and Parameters Setting}
\noindent In this study,  parking spaces on 420 streets containing 3905 parking spots in total in San Francisco are selected. The geolocation information of these on-street parking spaces is obtained from \cite{parking data san Fran}. We set $T$ to be 30 minutes. According to the on-street parking data set, the high parking activity period in San Francisco typically occurs between 6:00 and 19:00 (as illustrated in Fig. \ref{Fig.parking change}). Thus, we only consider this 13-hour period. The whole period is divided into 52 intervals, each lasting 15 minutes. According to Proposition 1, if at least one detection exists in every interval, the minimum required parking information updating frequency is satisfied. As a result, the total number of constraints in this problem amounts to $420\times52=21840$.

\begin{figure*}[!t] 
\centering 
\subfigure
{
    \begin{minipage}{1 \textwidth}
    \includegraphics[width=0.485\textwidth]{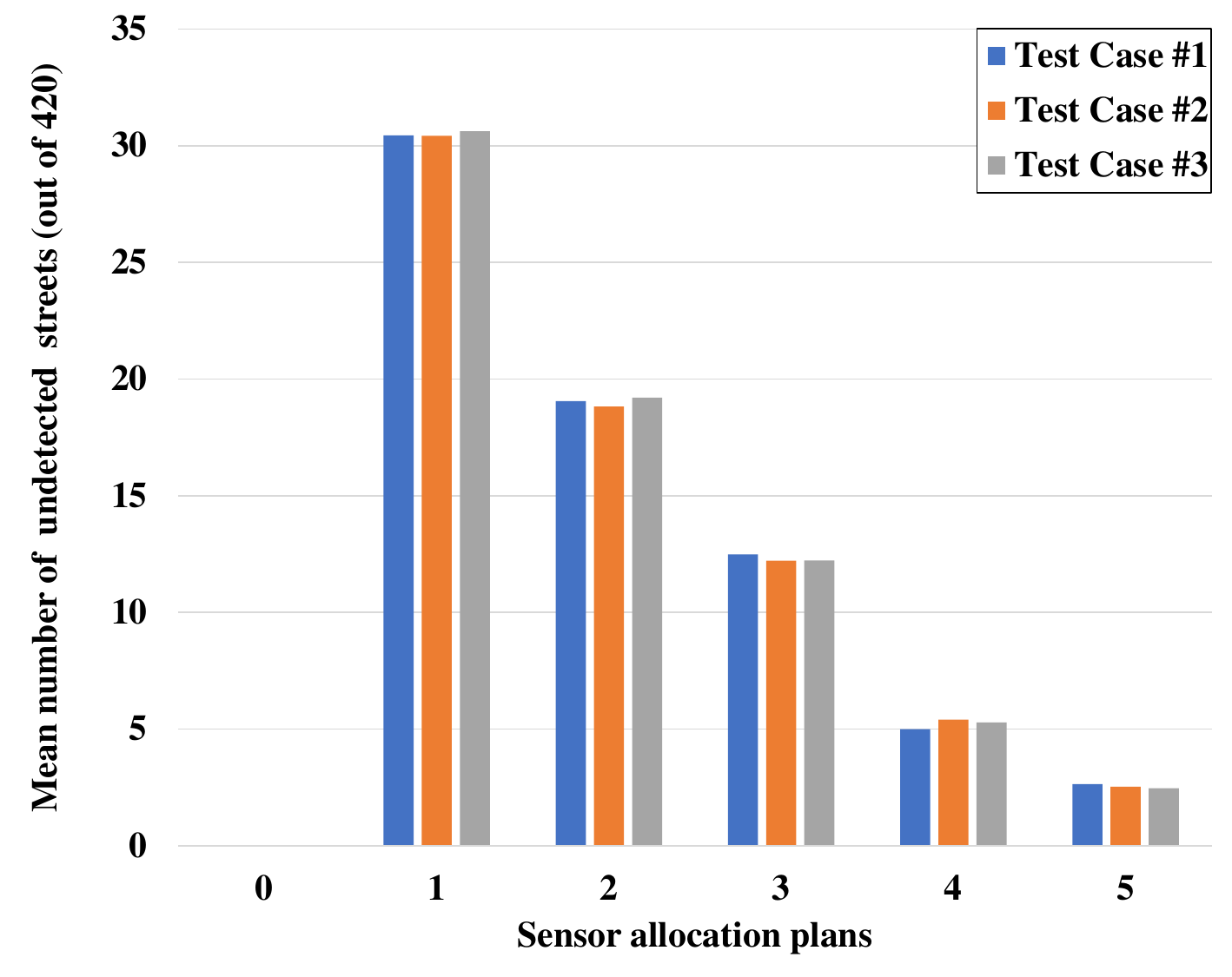}\vspace{4pt}
    \space \includegraphics[width=0.485\textwidth]{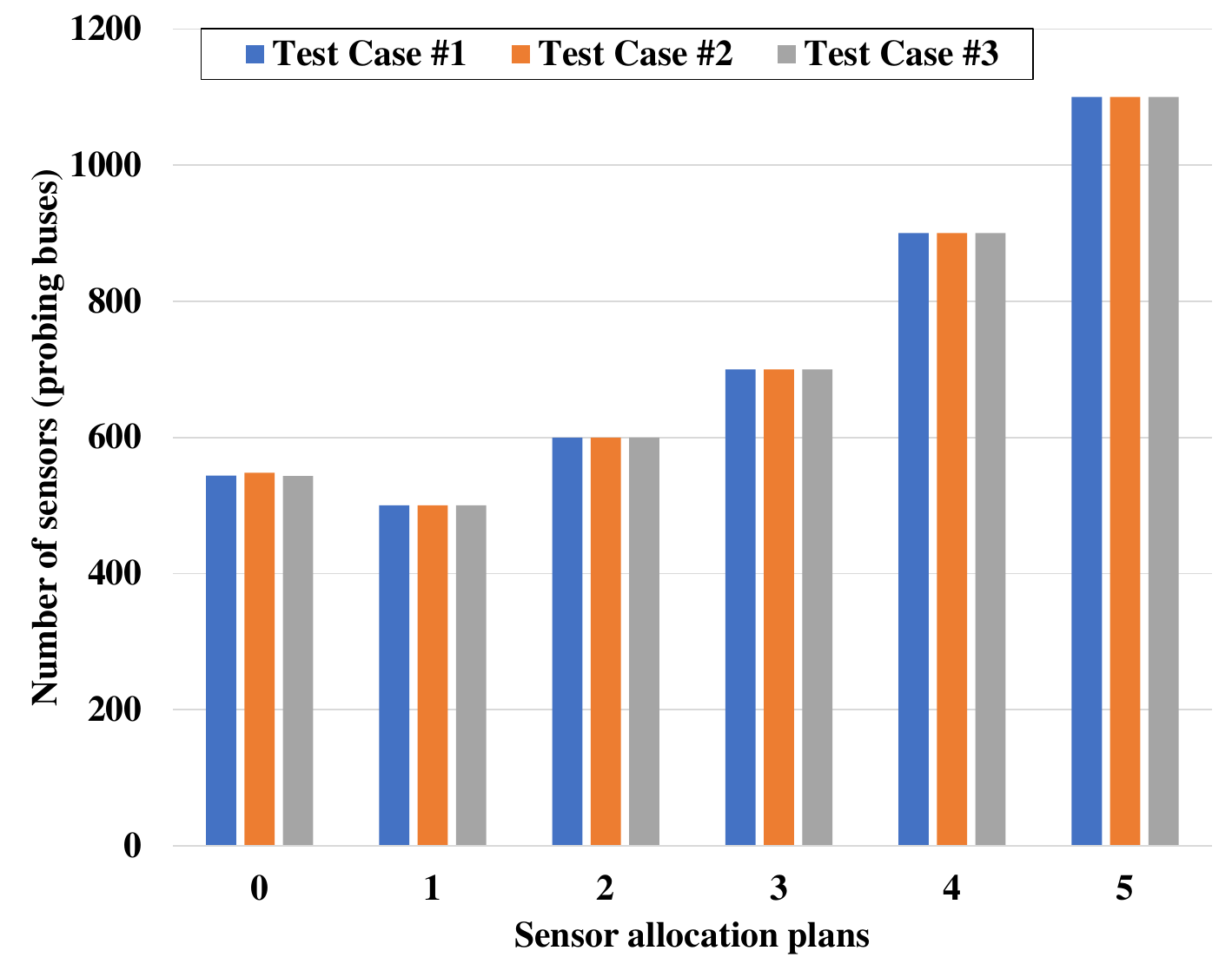}
    \end{minipage}
    
}
\caption{Average number of undetected streets per 30-minute time interval for various sensor allocation solutions} 
\label{Fig. results} 
\end{figure*}

\begin{figure*}[!b] 
\centering 
\subfigure
{
    \begin{minipage}{1
    \textwidth}
    \includegraphics[width=0.33\textwidth]{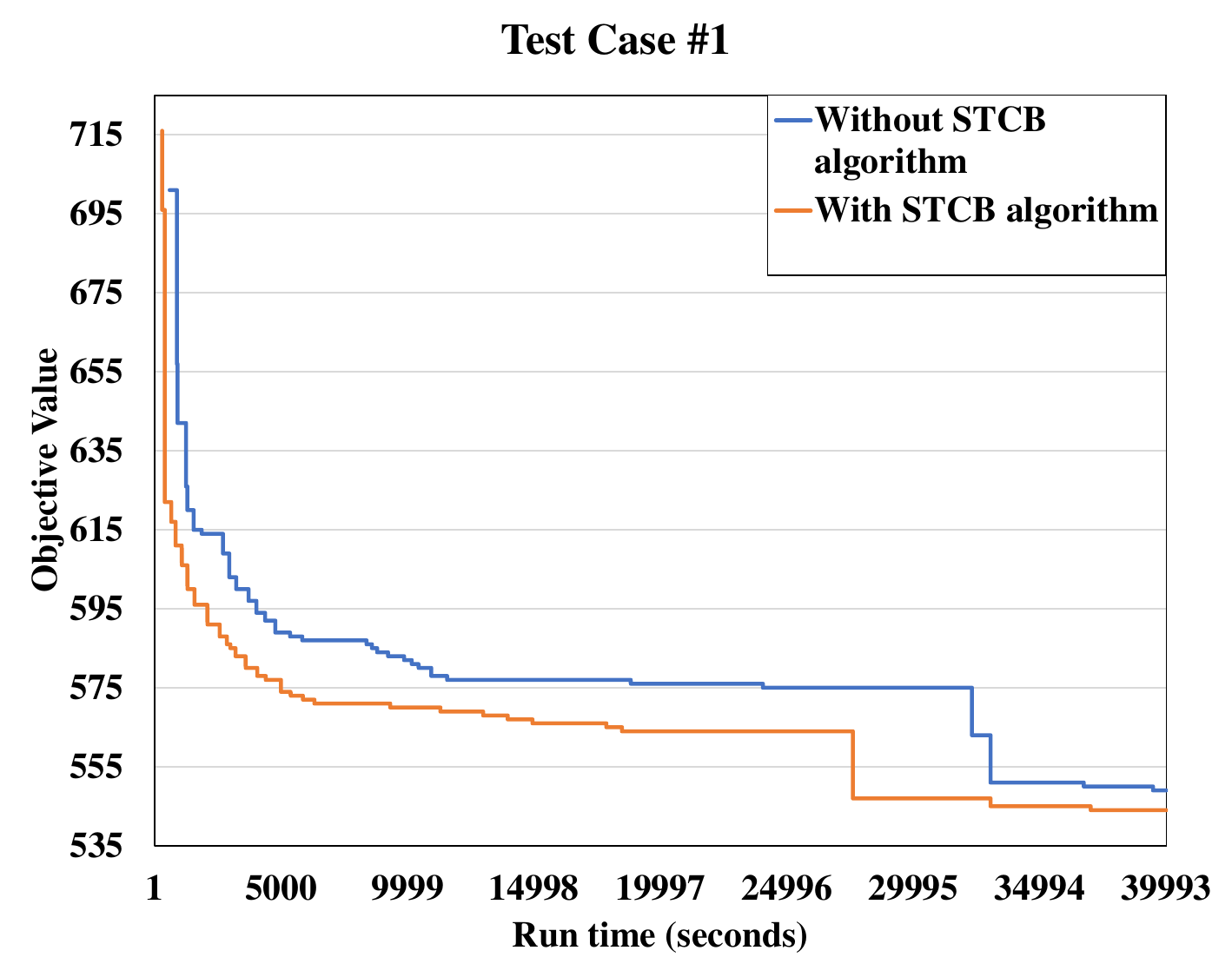}
    \includegraphics[width=0.33\textwidth]{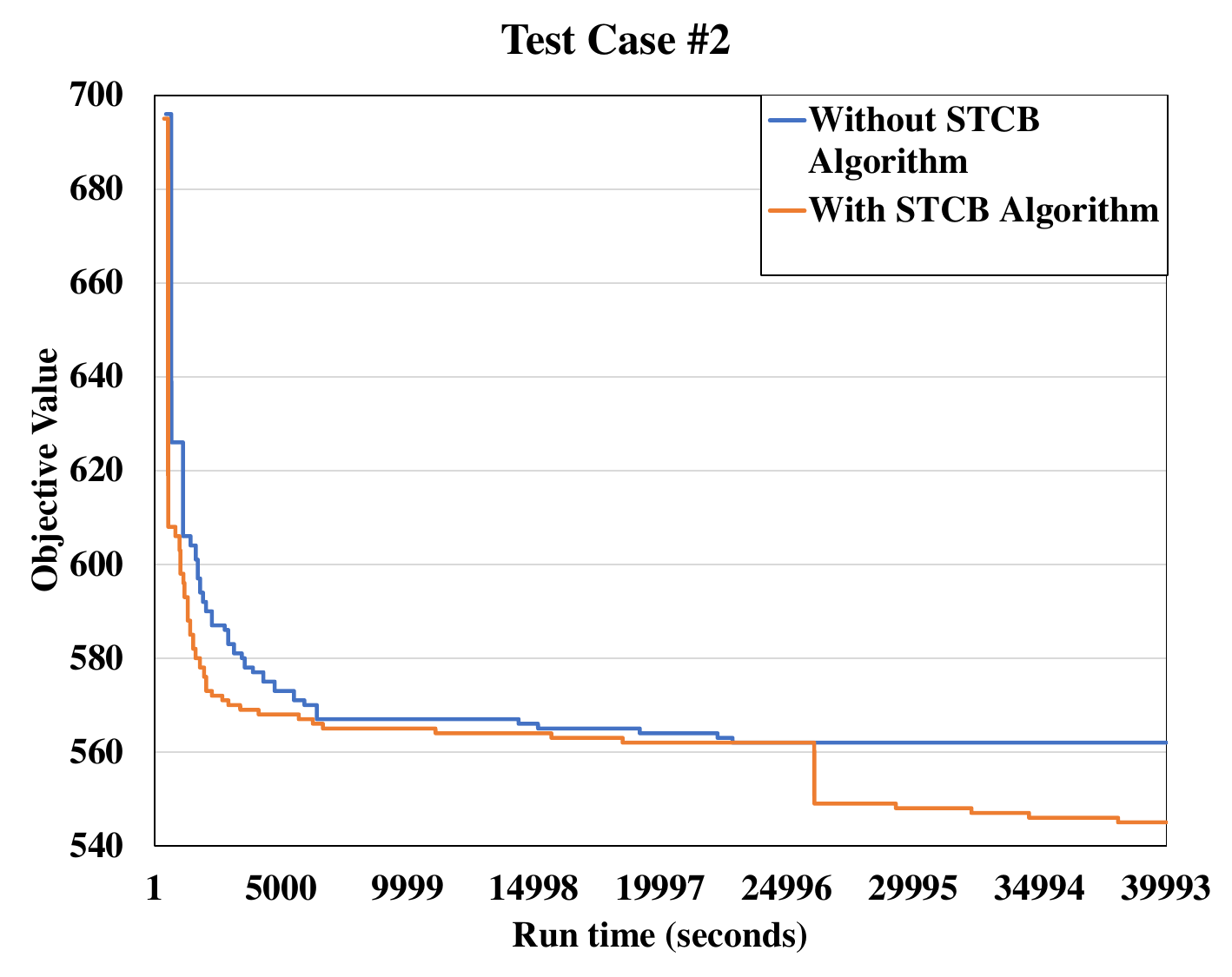}
    \includegraphics[width=0.33\textwidth]{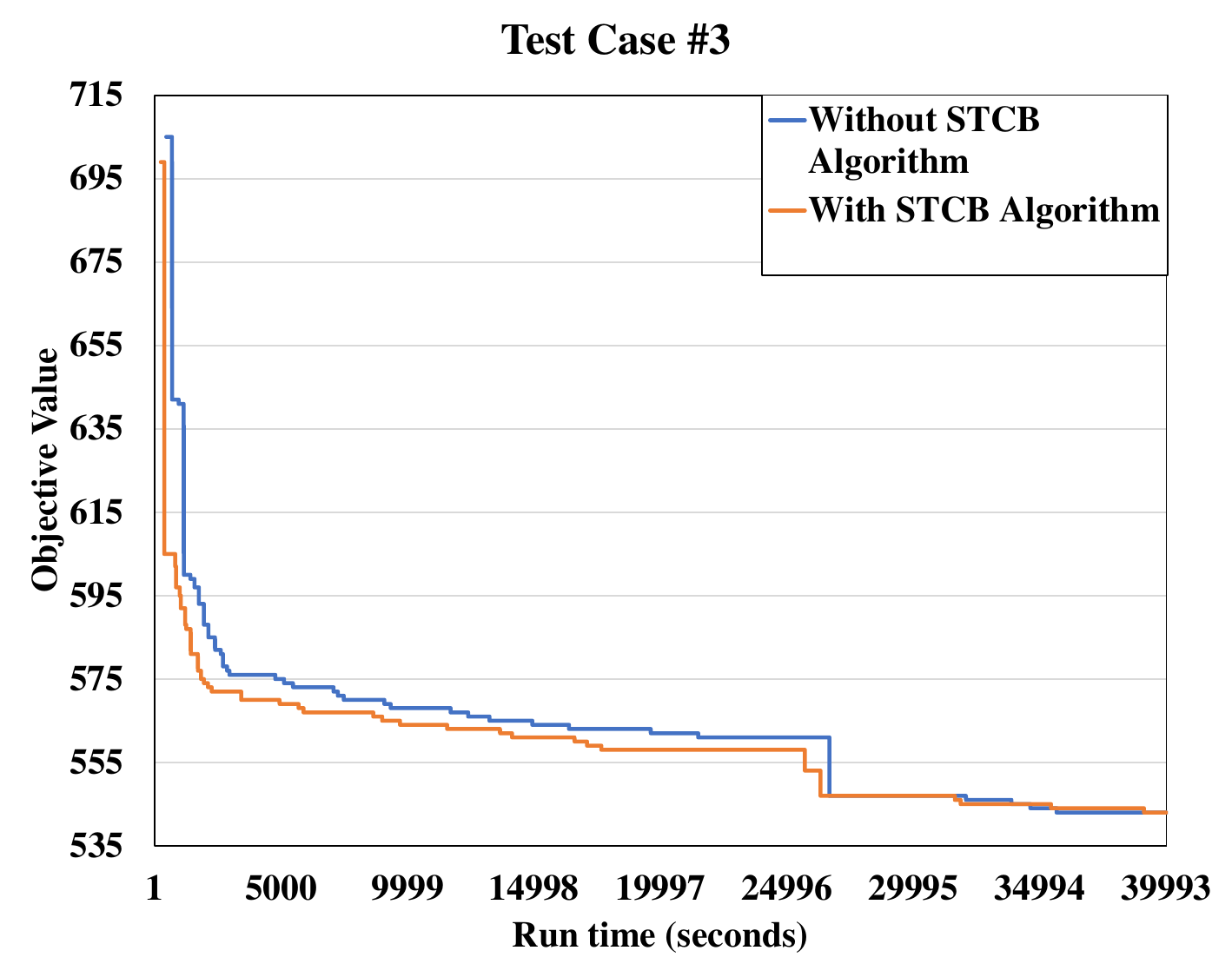}
    \end{minipage}
    
}
\caption{Performance comparisons between benchmark and STCB} 
\label{Fig. performance eval} 
\end{figure*}

We generate three random sets of bus schedules, with each  consisting of 400 routes. For every individual bus route, 12 buses are assigned, with their departure times staggered at 5-minute intervals (e.g., if the earliest bus departs at 7:00, the latest bus departs at 7:55). In other words, each generated set comprises $400\times12=4800$ buses, which corresponds to the number of decision (binary) variables in the optimization model.

The average cruising speed of a bus is set at 30 km/h. Given any one of the three bus route collections, the trajectory information for all buses can be estimated using the process introduced in Section \ref{subsec: prob cars}. This process generates the constraint matrix $\mathbf{A}$ for the optimization model. Numerical experiments are carried out using the previously established settings and three test cases, which are referred to as the bus route \#1, \#2, and \#3, respectively. Table~\ref{tab1} presents detailed information for the corresponding constraint matrices. The constraint matrices exhibit an average density of 0.58\%.

\begin{table}[!htbp]
\begin{center}
\caption{Information of Constraint Matrices}
\label{tab1}
\resizebox{0.48\textwidth}{!}{
\begin{tabular}{| c | c | c | c |}
\hline
Bus routes & Rows, & Average non-zero & Matrix \\
collection & columns & entries per row & density \\
\hline
\#1& 21840, 4800 & 27.76 & 0.578\% \\
\hline
\#2&21840, 4800 & 27.83 & 0.580\% \\ 
\hline
\#3&21840, 4800 & 27.69 & 0.577\% \\
\hline 
\end{tabular}
}
\end{center}
\end{table}

\begin{table*}[!b]
\centering 
\caption{Time comparisons of reaching different objective values}
\begin{minipage}[]{0.32\textwidth}
    \centering
    \captionsetup{labelformat=empty}
    \caption{Test Case \#1}
    \resizebox{1.05\textwidth}{!}{
    \begin{tabular}{c|c|c|c}
      \hline
     Obj. & Benchmark & STCB & Percent \\
     value &  runtime (s) & runtime &  speedup \\
      \hline
      622 & 1290 & 399 & 223.31\% \\
    606 & 2945 & 1070 & 175.23\% \\ 
    580 & 10431 & 3595 & 190.15\% \\
    574 & 32312 & 4990 & 547.54\% \\
    563	& $\mathbf{>}$40000 & 27603 & / \\
    \hline
    \end{tabular}
    }
  \end{minipage}
  \hfill
  \begin{minipage}[]{0.32\textwidth}
    \centering
    \captionsetup{labelformat=empty}
    \caption{Test Case \#2}
    \resizebox{1.05\textwidth}{!}{
    \begin{tabular}{c|c|c|c}
      \hline
     Obj. & Benchmark 	& STCB & Percent \\
     value &  runtime (s) & runtime &  speedup \\
      \hline
      608 & 1117& 529 & 111.15\% \\
580 & 3448 & 1616 & 113.37\% \\
572 & 5506 & 2259 & 143.74\% \\
565 & 15155 & 6647 & 128.00\% \\
549	& $\mathbf{>}$40000 & 26080 & / \\ 

    \hline
    \end{tabular}
    }
  \end{minipage}
  \hfill
  \begin{minipage}[]{0.32\textwidth}
    \centering
    \captionsetup{labelformat=empty}
    \caption{Test Case \#3}
    \resizebox{1.05\textwidth}{!}{
    \begin{tabular}{c|c|c|c}
      \hline
     Obj. & Benchmark 	& STCB & Percent \\
     value &  runtime (s)  & runtime &  speedup \\
      \hline
605 & 1147 & 375 & 205.87\% \\
588 & 1938 & 1200 & 61.50\% \\ 
572 & 7067 & 2247 & 214.51\% \\
567 & 11700 & 5884 & 98.84\% \\
560 & 26675 & 16598 & 60.71\% \\

    \hline
    \end{tabular}
    }
  \end{minipage}
\label{Fig. runtime comparison} 

\end{table*}

\subsection{Computational Results}

\noindent 
This section presents the computational results of the proposed model, with the objective of evaluating its ability to provide a high-accuracy and low-cost sensor allocation plan. It is important to note that the parking information update frequency of the crowdsensing system in \cite{mobile1} is heavily dependent on whether a probing bus passes an on-street parking spot within a short time frame (i.e., the duration of the gap between two consecutive detections). The detection system's accuracy diminishes when it fails to identify streets with parking spots in a specified period (i.e., no probing bus passes these streets within the time interval). Consequently, the experiment assesses the crowdsensing system's information accuracy by measuring the number of undetected streets in a 30-minute window and evaluates the cost based on the number of sensors (or probing buses) in use. 

As shown in Fig. \ref{Fig. results}, the  experiment utilizes the three bus routes, with their results represented in blue, orange, and grey, respectively. The left graph in Fig. \ref{Fig. results} depicts the average number of undetected streets (out of 420 streets) within a 30-minute period for various sensor allocation plans. Plan 0 corresponds to the optimal (minimal sensors) allocation policy as determined by the proposed model. For all three test cases, the optimal solutions require around 550 probing buses for the 420 streets (with 3905 parking spots in total). The remaining plans, i.e., Plans 1-5, serve as control groups that randomly select a fixed number of buses (500, 600, 700, 900, and 1100, respectively). The number of buses employed for these plans is displayed in the right graph of Fig. \ref{Fig. results}.

Apart from the deterministic policy proposed by Plan 0, each of the remaining plans is tested 10 times to mitigate variances, as these plans employ random policies. The results demonstrate that, when utilizing randomized plans, low undetected rates (or high accuracy) are achieved only when more than 1100 buses are used as probing vehicles. Conversely, when implementing the deterministic plan provided by the optimization model, only approximately 550 buses are needed, which is roughly half the cost of the aforementioned randomized plan. This outcome validates that the proposed model can ensure the maximum limit of the time gap between two consecutive detections for crowdsensing systems while significantly reducing the total cost.

Furthermore, it can be observed from Fig. \ref{Fig. results} that the differences between the three test cases are not substantial, suggesting that the model proposed in this paper is not sensitive to different bus route configurations.

\subsection{Performance Evaluation of STCB}
\noindent This section presents a comparison of the speed results when solving problems with and without the STCB algorithm. Figure \ref{Fig. performance eval} illustrates the general run-time performance for the three test cases (three bus routes), where the blue line represents the results of direct solving using the Gurobi solver, and the orange line denotes the results obtained with the STCB algorithm. Since the goal is to minimize the total number of probing vehicles, a lower objective value indicates a better outcome. 

To obtain a more precise  understanding of the algorithm's acceleration effect, Table \ref{Fig. runtime comparison} presents the time (in seconds) needed to achieve  representative objective values using both direct solving and the STCB algorithm. The percentage of speedup is calculated using the formula $\mathbf{\frac{Benchmark \ runtime}{STCB \ runtime}-100\%}$. It can be observed that for most objective values presented in the table, the STCB algorithm provides a speedup of more than 100\%. Notably, after solving for over 15,000 seconds, the STCB algorithm in all three test cases saved more than 10,000 seconds. These results demonstrate that the STCB algorithm can substantially enhance the solving speed of the proposed model.

\section{Conclusions and Future Work}
\label{conclu}
\noindent This paper presents a general framework for allocating sensors in a crowdsensing system for on-street parking occupancy. Through theoretical proofs and numerical tests, it is demonstrated that the proposed mathematical model can provide an optimal allocation plan that minimizes the total number of sensors while maintaining detection accuracy or  frequency. Additionally, given the computational complexity of solving the mathematical model, a self-trained cardinality-branching (STCB) algorithm is designed to accelerate the solving process. The performance comparison indicates that the STCB algorithm effectively enhances the model-solving speed.


Flexibility entails the consideration of additional factors, such as varying parking activities. In practical applications, different regions typically exhibit distinct activity levels, which may fluctuate during specific time intervals. For example, parking changes tend to be more frequent around 6-8 AM and 5-7 PM, while remaining relatively stable between 8 AM and 5 PM and less frequent during other time intervals. To enhance the model's flexibility, future work may consider modifying the coverage constraints of the optimization model. One potential approach is to subdivide time intervals based on parking change activity levels and adjust the constraints to cover all these sub-intervals.

\section*{Acknowledgement}
\noindent The authors would like to sincerely express their gratitude to Sun Yan from the Chinese University of Hong Kong, Shenzhen for providing valuable insights on fine-tuning the hyper-parameters of STCB.

\end{document}